\newcommand{\beq} {\begin{eqnarray}}
\newcommand{\eeq} {\end{eqnarray}}
\documentclass[12pt]{JHEP3}

\title
{Nothing for Branes}
\author{Daniel Green \\ SLAC and Department of Physics, Stanford University, Stanford, CA 94305-4060 \\ \\KITP, University of California, Santa Barbara, CA 93106-4030 \\ Email: \email{drgreen@stanford.edu}}

\preprint{SU-ITP-06/30 \\ SLAC-PUB-12179}

\abstract{Recent work on closed string tachyon condensation suggests the existence of a `nothing state' where closed strings and space itself vanish.  We consider the evolution of D-branes in such backgrounds, focusing on the early stages of the condensation process.  We find evidence that the branes exist in the region; although, generically their apparent mass grows exponentially with time.  However, there exist specific branes whose boundary states are unaltered by the tachyon.}
\begin{document}
\section{Introduction}

It has long been speculated that closed string tachyons lead to the decay of spacetime.  For spatially localized tachyons, it has been clear for several years that this picture is correct and gives a decay from one space(time) into another (e.g. \cite{Adams:2001sv, Adams:2005rb, Headrick:2004hz, Martinec:2002tz}).  Recent work has provided insights into decays in non-localized cases \cite{McGreevy:2005ci, Yang:2005rx, Hikida:2005xa, Nakayama:2006gt, She:2005mt, Berkooz:2005ym}.  In particular, from a worldsheet perspective, the condensation process gives closed string modes exponentially growing masses.  This is consistent with the idea that there are no closed string states when the tachyon has fully condensed.

This suggests that there is a vacuum state of the string configuration space that contains nothing, meaning neither matter nor space-time.  However, there is much more to the configuration space of string theory than perturbative closed strings.  One might wonder if there can be excitations of this state other than the spacetime from which we condensed, namely D-branes without a spacetime interpretation.

This problem has been studied very little for good reason.  In the absence of a worldsheet or spacetime description, there are few tools left with which to study the brane evolution.  The results from the $c=1$ matrix model have provided hints at an answer \cite{Karczmarek:2004ph}.  In 2d string theory, closed strings in the bulk are given by the collective excitations of the fermi sea formed from free fermions in an inverted quadratic potential.  Meanwhile, an individual fermion describes the worldvolume theory of a single D0 brane.  When tachyon condensation occurs, the fermi sea drains, thus closed string excitations vanish in the absence of any collective phenomena.  However, one could still have extra fermions separated from the sea that survive when the sea drains.  As a result, there is a clear difference in the matrix model when branes are present in `no-man's land'.

While these results are interesting, there remain a few questions.  The branes that appear in the matrix model (the free fermions) are unstable, not only in the bosonic string but also in the type 0A and 0B models ($\hat{c}=1$).  One might like to know if stable branes can survive in this region as well.  In addition, the `tachyon' in $c=1$ is massless; thus, the condensation is added by hand.  It would be helpful if we could address this question in the context of stable branes in string theories where the tachyon must condense.

On the other hand, work on localized closed string tachyons has provided evidence that seems contrary to that from $c=1$.  In particular, in the study of D0 branes on $\mathbb{C} / \mathbb{Z}_N$ \cite{Adams:2001sv}, the D0 branes at the tip gain a large mass.  The open string potential is modified by the tachyon background in such a way that the D-branes move out of the region leaving a smoothed-out space.  This would seem to suggest that branes are lifted in the tachyon condensate in much the same way as the closed string states.

We will examine these questions in the context of the `Nothing' state presented in \cite{McGreevy:2005ci}.  The first reason to consider this scenario is that all D-branes will see the tachyon because the condensate is not spatially localized.  In this case, only spacelike branes can survive by simply by avoiding the region where the condensate lies.  The second reason is simply that the worldsheet theory is very similar to Liouville theory.  For this reason, we can focus on the aspects of the worldsheet description that also appear in the non-perturbative description of the $d \leq 2$ string.  In particular, we will discuss the boundary state and the one-point function on the disk which describe the branes in terms of the closed strings they source.  In order to get intuition for these quantities, we will relate the disk one point function to the one point function in a field theory model with an arbitrary source.  This will help separate out the effect of the tachyon on the brane from its effect on the closed strings the brane sources.

The results presented here suggest that branes generically gain large masses in the tachyon background.  There are branes whose boundary states seem insensitive to the tachyon background; although they will gain a large effective mass from the closed strings they source.  These results arise in a way consistent with all previous results and suggest the origin of the seeming discrepancy in earlier work.

The paper is organized as follows: in section 2 we will review the worldsheet RG description of tachyon condensation, focusing on the model in \cite{McGreevy:2005ci}. In section 3, we will consider the same model on a worldsheet with boundaries.  The behavior of the branes will be inferred from the boundary state description at early times when the tachyon vacuum expectation value is small.  In section 4, we will discuss the interpretation of these states and we will discuss how this approach fits with other results.  In section 5 we will make the connection with conformal perturbation theory.  Finally, in section 6, we will discuss some interesting open questions and future directions of study related to these results.

\section{Tachyons and the Worldsheet}

In \cite{McGreevy:2005ci}, a Milne universe in the critical dimension was considered where a circle was given a time dependent radius and anti-periodic boundary conditions on the fermions.  The remaining eight dimensions were left flat.  This example is characteristic of a larger class of cosmologies which were also considered, but we will focus on the Milne case for concreteness.
  
At early (or late) times, when the radius of the circle is string scale, a winding tachyon appears in the spectrum.  In a type II theory, the worldsheet theory in the presence of the condensing tachyon is deformed by the operator
\beq
\label{tachyon}
\int d^2\theta (2 i \mu e^{-\kappa \tilde{X_{0}}} \cos(w \tilde{\bar{\Omega}}) ),
\eeq
where $\bar{\Omega}$ is the T-dual coordinate on the circle and the tilde denotes the superfields where $X_0$ and $\Omega$ are the bosonic components.  If we integrate the fermionic coordinate, the bosonic action is
\beq
\label{action}
iS_{bos} = i \int d\tau d\sigma( (-\partial X_0 \bar{\partial} X_0) + v^2 X_0^2 \partial \Omega \bar{\partial} \Omega +\partial X_i \bar{\partial} X_i \nonumber \\
+{\mu^2} e^{-2\kappa X_0} \{ (\frac{w}{v X_0}\sin(w {\bar{\Omega}}))^2-(\kappa \cos(w {\bar{\Omega}}))^2 \}).
\eeq
Where $X_{i}$ are the flat eight dimensional coordinates and $v$ is the rate of expansion of the universe.

The exponentially growing term acts like a mass that should suppress the path integral when $X_0 \ll -1$.  This was demonstrated in \cite{McGreevy:2005ci} when using a Wick rotation on all the spatial coordinates in order to keep the term $e^{-\kappa X_0}$ real.  After rotation, the integral is convergent and it was shown in \cite{McGreevy:2005ci} that by integrating out the zero mode of $X_0$ the support of the path integral lies only in the region without the tachyon.  The choice of Wick rotation of the worldsheet theory is equivalent to a choice of state of the spacetime theory.  In general, the analytic properties of the configuration space of string theory are largely unknown.  In this case, the results seem insensitive to the choice, but this is not generally true of time dependant backgrounds as shown by \cite{Nakayama:2006qm}.

In our analysis, we will take the more standard rotation,
\beq
\tau \equiv e^{-i\gamma} \tau_{\gamma}, X_{0}\equiv e^{i\gamma} X_{0, \gamma}, \kappa \equiv e^{-i\gamma} \kappa_{\gamma}, v \equiv e^{-i\gamma} v_{\gamma}.
\eeq
We would like to be able to use the tools of conformal perturbation theory as much as possible, and thus it would seem pertinent to maintain semi-classical conformality.  One approach to this problem is to add a background charge in the Euclidean theory to make up the difference (coefficient of the dilaton coupling).  In this case, we will need a dilaton coupling of the form $2\kappa R X_0$.  In general, these are subtle issues whose validity we will not address.  Nevertheless, our analysis does not require this particular rotation in a crucial way, thus it seems unlikely that qualitative features we will discuss are at risk.

In particular, this analysis applies under two assumptions.  First of all, we will work only in the regime where the semiclassical action is a good description of the dynamics (namely (\ref{action}) ).  This will be true when the tachyon vacuum expectation value is small and can be treated as a perturbation.  The size of the vev and time derivatives are controlled in part by $\mu$ and $\kappa$ which will be taken as small.  Everything will be treated only to leading order in these parameters.  This will be valid at early times (as measured compared to $\kappa$ and $\log(\mu)$), but will break down well before the $X_0 \rightarrow \infty$ limit.  Secondly, we will ignore the contributions of the running dilaton except for its effect on the scaling dimensions of operators.  The dilaton will be involved at later times as discussed in \cite{Freedman:2005wx}.

\section{Branes and Boundary states}
In this section, we will examine D-branes in these closed string tachyon backgrounds using their worldsheet description.  We will ultimately be interested in branes that lie in the `Nothing' region described in \cite{McGreevy:2005ci} in the type II description.  On the worldsheet, branes will appear as boundaries and will generically break supersymmetry.  In order to preserve some of the supersymmetry, boundary terms need to be added to the action.  In particular, given an action of the from
\beq
\int d^2z d^2\theta \mathcal{L}(\theta,\bar{\theta}) ,
\eeq
one should add a boundary term of the form \cite{Douglas:2003up}
\beq
i \eta \oint dx  \mathcal{L}(\theta = \bar{\theta}=0).
\eeq
The supersymmetry variation of the boundary action will cancel the boundary terms from the bulk variation with respect to $Q + \eta \bar{Q}$ where $\eta=\pm 1$ and $Q$, $\bar{Q}$ are the worldsheet SUSY generators.  In the case under consideration, using the action from (\ref{action}), we need to add the boundary action 
\beq
\mathcal{L}_{bdy} = -\eta \oint dx (i\psi_{\mu} \bar{\psi}^{\mu} - 2\mu \cos(w \bar{\Omega}) e^{-\kappa X_0}).
\eeq
It will be useful to think in terms of a T-dual picture where $\bar{\Omega}$ is a coordinate on the circle.  In particular, we will take Dirichlet boundary conditions on T-dual coordinate such that we can specify the value of the cosine term.  These correspond to wrapped branes with a Wilson line.  In this case, we will be able to decouple the $\Omega$ boundary conditions from the $X_0$ boundary conditions.  The D0 branes are a more difficult problem in this setup and we will not address it here.  See \cite{Adams:2001sv} for results on D0s in a slightly different context.

The variation of the action with respect to $X_0$ produces the boundary terms,
\beq
\label{bc}
\oint dx 2 \delta X_0 (\frac{\partial X_0}{\partial y}-\eta \kappa \mu \cos(w \bar{\Omega})e^{-\kappa X_0}).
\eeq
The Neumann condition takes a form similar to that of an FZZT brane\cite{Fateev:2000ik}.  If we build a boundary state with an $\bar{\Omega}$ eigenfunction, we can replace the operator in the $X_0$ boundary condition with its eigenvalue.  Furthermore, because the second term added to the boundary action is bosonic, the fermion boundary conditions are those from flat space.  

At this point we will not demand conformal invariance beyond the classical level.  These boundary conditions are classically conformal, but not all will be conformal at higher order in $\alpha^{\prime}$.  We will return to discuss these points in section 5.  In any case, the bulk perturbation is not exactly conformal so we will proceed via conformal perturbation theory.

The minisuperspace wavefunction for the wrapped brane boundary condition is
\beq
\label{bwavefunction}
\Psi_{b}(X_0) = \exp(-\eta\mu e^{-\kappa X_0}\cos(w \bar{\Omega})).
\eeq
This is simply the zero eigenfunction of the minisuperspace quantization of the boundary conditions.  Stated differently, it is boundary state overlap with the $X_0$ zero-mode eigenfunctions.  From this basic calculation, we have a potentially interesting result.  This wavefunction takes the form familiar from Liouville theory; however, now we have both growing and decaying solutions (depending on the sign of $\eta\mu\cos(w \bar{\Omega})$).  Furthermore, stable branes in type II will ultimately be the superposition of both decaying and growing modes.  Our intuition from Liouville theory suggests these branes live in the `nothing state'.  In particular, in $c \leq 1$, one typically uses the point where $\mu e^{-\kappa \phi} \approx 1$ as representing the end of the brane.  We will not assume that conclusion since it is unclear it is appropriate in light of the growing solutions.  We will return to the subject of interpretations in the next section.  First, let us work out two representative examples.

\subsection{Branes at $w \bar{\Omega}= n\pi$}

First, we will consider the wrapped brane with $w \bar{\Omega}= n\pi$ where $n$ is an integer.  The one-point function contains all the information about the boundary state, but it has a more direct analog in field theory which we will use for interpretation in section 4.  For this reason, we would like to calculate the one-point function in the CFT for later use.  For the wrapped branes, the boundary state factorizes, and we can deal with the $X_0$ and $\bar{\Omega}$ parts independently.  For this choice, it is relatively easy to treat the $\bar{\Omega}$ direction using the method of images.  However, the boundary conditions on $X_0$ are non-trivial and it is not clear how they will contribute.

In order to make an estimate of the $X_0$ component, we will use a minisuperspace truncation on the effective theory in the $X_0$ direction.  From the wavefunctions for the bulk modes, the one-point function is given by
\beq
\label{inner}
\langle p | B \rangle = \int dX_{\mu} d\psi_{\mu} d\bar{\psi}_{\mu} \Psi^{\dag}_p \Psi_{B}.
\eeq
where $| B \rangle$ is the boundary state and $\Psi_p$ is the wavefunction of a bulk mode with momentum $p$.  To first order in conformal perturbation theory, the $X_0$ direction behaves as
\beq
\label{Actionredux}
\int d^2z (\partial X_0 \bar{\partial} X_0 + \kappa^2 \mu^2 e^{-2\kappa X_0})+ S_{SUSY} -2\eta\mu\oint dx e^{-\kappa X_0},
\eeq
where $S_{SUSY}$ contains the fermions related by supersymmetry (where we have absorbed any $\Omega$ normalization subtleties into $\mu$).  Because we are dealing with Dirichlet boundary conditions on $\Omega$, $\Psi_b \propto \delta(\bar{\Omega}-\bar{\Omega}_0)$ (for the full boundary state, not only the minisuperspace wavefunction).  For this reason, the contributions for the $\Omega$ fields will be unimportant for the qualitative behaviour.

A very similar calculation is done in \cite{Douglas:2003up} where the minisuperspace mode expansion was calculated explicitly.  If we impose the boundary conditions that the modes decay under the effective `wall' then in the NS sector
\beq
\Psi_p \propto K_{ip}(\mu e^{-\kappa X_0})
\eeq
where $K_{\alpha}$ is the modified Bessel function of the second kind.  The R sector wavefunctions are superpositions of $K_{ip \pm {1 \over 2}}$.  Asymptotically, these behave as
\beq
\Psi_p \approx \exp(-\mu e^{-\kappa X_0}).
\eeq 
If we combine this with the boundary wavefunction (\ref{bwavefunction}) with $\bar{\Omega}=0$ and $\eta=-1$, then the two exponentials combine in (\ref{inner}) to give a constant.  Thus, the one-point function finds support arbitrarily deep in the condensate.

We could import the exact boundary Super-Liouville results, but it is unclear the solutions should agree at later times or at higher orders in conformal perturbation theory.  The above results agree with the semi-classical limit of the exact super-Liouville solutions.  Therefore, the differences between the approach taken here and using exact Liouville solutions are significant outside the range of validity of our approximation.  The minisuperspace result is also organized in a way that makes interpretation easier.  Exact solutions to boundary (super)sine-Liouville might prove useful but have not been calculated to date, as far as we know.

Recalling that any wrapped brane will be described by $\Psi_B = \Psi(\eta=+1)\pm\Psi(\eta=-1)$, all D-branes (expect for those with $w\tilde{\Omega} = {\pi \over 2} + n \pi$) will have an exponentially growing piece.  In this treatment, the branes continue into this region, unlike the perturbative strings.  This is a choice of state, which ideally could be chosen in the same manner as the closed strings.  We will return to the issue of interpretation in section 4.

\subsection{Branes at $w\bar{\Omega} = {\pi \over 2} + n \pi$}

As suggested by the previous subsection, the case where $w\bar{\Omega} = {\pi \over 2} + n \pi$ seems qualitatively different from the more generic boundary conditions.  What makes this particular choice interesting is that the boundary term vanishes.  The significance of this is simply that $\cos(w\bar{\Omega})|B\rangle = 0$, so all the remaining boundary conditions are unchanged by the boundary term.  To be explicit, the boundary state has the property
\beq
|B,w\bar{\Omega} = {\pi \over 2}, \mu \rangle \equiv |B,w\bar{\Omega} = {\pi \over 2}, \mu=0 \rangle.
\eeq
As a result, the boundary state for this particular brane is unaltered by the bulk tachyon.  In fact, if one ignored the time dependence from the shrinking radius of the circle, this boundary state can be explicitly constructed.  For this reason, we expect that the brane survives in the tachyon condensate and its `bare' mass is unaltered.  However, because these branes source fields their effective mass should still grow rapidly as the closed strings are massed up.

By comparison with the last section, it should be clear that the one-point function is significantly altered by the tachyon, even though the boundary state is unchanged.  In particular, the string propagator is still exponentially suppressed.

\subsection{Caveats}

One might also be concerned about the divergence of the wavefunction as $X_0 \rightarrow -\infty$.  In particular, we could worry that these states are non-normalizable.  We will not address this directly, but it should be pointed out that this analysis is only valid in the weakly coupled, small tachyon regime, which is far from the $X_0 \rightarrow -\infty$ limit. Thus, the domain of validity of the worldsheet action is a finite interval in time and thus there are no problems with non-normalizability.   Furthermore, when we Wick rotate back to Lorentzian signature, these modes will become oscillatory.  It is also possible that the dilaton could roll to strong coupling in this process \cite{Freedman:2005wx}, which will alter the problem dramatically.  It remains unclear whether this will actually be a problematic limit and we will not address it here.  In 2d type 0 theories one might expect this problem. The extended branes of both types are present in type 0 but there is an open string tachyon on their world volume.  This additional term ensures that, for suitable choices of the boundary cosmological constant, the wavefunction is normalizable.

One might also wonder what role these branes play in the context of black hole evaporation.  In \cite{Horowitz:2006mr}, a family of final states were found for the black hole in which unitarity was maintained.  Additional states with branes could be non-unitary.  There are a number of reasons that this is likely not to be a problem.  First of all, the branes all generically get mass (as we will see), either from the boundary state itself (bare) or from the strings it sources (dressed).  Ultimately, one should be able to choose a final state without massive branes as is done for the massive closed strings.  Secondly, the branes being discussed here are wrapped on the circle.  Therefore, they have measurable charge and have a significant mass set by the size of the circle at infinity.  Thus these branes have a very classical role in the system and thus are difficult to incorporate in a scheme where the evolution is non-unitary.

A remaining question is simply whether it is possible to construct a brane anti-brane system such that the branes survive without developing an open string tachyon.  While it is true that the two can reach the condensate at different times, the open string tachyon may still develop inside the condensate.  For example, the ZZ brane lies at infinity in the strong coupling region of Liouville theory (which is equivalent to a spacelike brane at $X_0 = -\infty$ in this model).  However, the closed string one point function is non-zero and the open string spectrum contains two states (a tachyon and a massless gauge field).  All of these branes eventually decay despite the presence of a background that one might naively expect would lift the mass of the tachyon.

\section{Interpretation}

\subsection{Growing Sources}
Because we are studying the source of closed strings in the presence of a tachyon background, it is hard to distinguish between the effects from the string and the effects from the brane.  In an attempt to separate these two independent contributions, let us consider a field theory model of the process.
We will take a scalar field $\phi$ with a mass that grows exponentially in time, but is given some arbitrary source $j(t,x)$.  The scalar field will play the role of the closed strings and the source will be a D-brane.  The Lagrangian for this system, by analogy with the previous section is
\beq
\mathcal{L} = \partial_{\mu} \phi \partial^{\mu} \phi-\mu^2\kappa^2 e^{2\kappa t}\phi^2+j(t,x)\phi.
\eeq
We are interested in how the one point function $\langle \phi \rangle = \varphi$ is related to $j$.  In this model, the one-point function is simply the solution to the equations of motion.  We will proceed in the Euclidian theory, but the relation to the Lorentzian theory will be completely transparent.  We will work in momentum space in the `spatial' directions for simplicity.  Thus, we are looking for the solution to
\beq
\partial_{t}^{2} \varphi - \mu^2\kappa^2 e^{2 \kappa t}\varphi-p^2\varphi = j(t,p).
\eeq
By the change of variables $z=\mu e^{\kappa t}$, this takes the form
\beq
\partial_z(z \partial_z \varphi) - (z+p^2z^{-1}) \varphi = j(z,p)z^{-1} = \tilde{j}.
\eeq
Because this is in Sturm-Liouville form, the differential operator is self-adjoint and hence the Green's function is given by
\beq
\partial_z(z \partial_z G(z,y)) - (z+p^2 z^{-1}) G(z,y) = \delta(z-y).
\eeq
The homogenous solution to this is simply the combination of modified Bessel functions (for the Lorentzian case one simply removes the word `modified').  Finding the Green's function simply requires taking solution from $z<y$ and $z>y$ and matching with the conditions that $G_+(z=y)=G_-(z=y)$ and $y(\partial_z G_+(z=y)-\partial_z G_-(z=y))=1$.

However, as usual, one is left with choice of boundary conditions on the Green's function which is the same as choosing the appropriate propagator.  As in the string theory example, we will define this by the Euclidian theory where there is a natural choice.  In particular, the Euclidian theory has exponentially growing and decaying solutions.  Thus, we require that the Green's function is finite as $t \rightarrow \pm \infty$.  This choice yields the solution
\beq
G(z,y) = K_{p}(z)I_{p}(y), \hspace{5mm} y<z \nonumber \\
= I_{p}(z)K_{p}(y),    \hspace{5mm} y>z
\eeq
If we take $p^2 \simeq 0$, and look at the asymptotic behaviour we expect that for $t \gg \tilde{t}$,
\beq
\varphi(t) \simeq \int d\tilde{t} j(\tilde{t}) \exp(-\mu |e^{\kappa t}-e^{\kappa \tilde{t}}|).
\eeq
In the Euclidian calculation, one sees these rapidly decaying exponentials for any source.

This is the same expression as the semiclassical one-point function (\ref{inner}).  In particular, the NS sector semiclassical mode is a modified Bessel function as well.  Thus, in this analogy $\Psi_B \rightarrow j$.  If we measure mass by coupling to gravitons, then the mass grows very rapidly.  By comparison, the flat space branes and the surviving branes in this picture simply have $\Psi_B \propto 1$, as one would expect for an object of constant tension.

To make a more direct appeal to the boundary state, for a brane in flat space, the tension of the brane arises in that context from the normalization of the state.  These branes only have delta function dependence on the zero modes for the Dirichlet direction and are independent of the zero modes in the Neumann directions.  Therefore, if we were to group the zero mode dependence into the normalization, these exponential contributions would take precisely the form of a time dependant tension/mass.

It should be pointed out that the mass growing as $m \propto e^{\mu e^{\kappa t}}$ may not be correct at later times (although this scaling could potentially be explained using methods from \cite{Craps:2006gk}).  This work is valid when the tachyon vev is small.  Hence, this is a good description to leading order in the perturbative expansion leaving
\beq
m \propto 1 + \eta \mu \kappa \cos(w \tilde{\Omega}) e^{-\kappa X_0} + \dots
\eeq
We can see that the choices of $\eta$ leave a growing and shrinking mass as the tachyon grows; however, the precise form the growth remains to be understood.

Finally, this interpretation, as stated before, applies only to the `bare' mass.  There are additional contributions to the total mass that come from the fields sourced by the brane.  Because these fields have exponentially growing masses, the `dressed' brane will be very massive \footnote{We would like to thank Eva Silverstein for stressing this point}.

\subsection{Comparisons to Previous Results}

D-branes in the presence of closed string tachyons was one of the major developments in the subject \cite{Adams:2001sv}.  There, the worldvolume theory of a point-like brane was studied as a closed string tachyon appears near the tip of the cone.  Before the appearance of the tachyon, the moduli space of the brane is precisely $\mathbb{C}/ \mathbb{Z}_N$.  The tachyon vev generically breaks the symmetries of the worldvolume theory, smoothing out the moduli space, and removing the tip of the cone.

However, if we demand that the D-brane remain at the tip, the energy of the solution (and hence the mass of the brane) grows as $T^2$ where $T$ is the tachyon vev.  This again is simply the leading order contribution to the potential and is only valid for small $T$.  Thus there is nothing inconsistent with the form of the growing masses in these two cases.

The study of D-branes in c=1 matrix model\cite{Karczmarek:2004ph} would seem to be slightly at odds with the above results.  In that context, branes survive as individual fermions far removed from the draining sea.  In particular, the claim there is that the set of branes that survive is certainly more than measure zero.  However, in that context, one is studying the bosonic string in two dimensions.  This is clear that the behaviour of the branes in this background is due entirely to the requirement of worldsheet supersymmetry and is thus not at odds with their results.

When considering unstable branes like those in bosonic theories, there are relevant operators allowed on the boundary (the open-string tachyon).  These operators lead to the decay of the brane, but can be tuned in order to have the decay occur after the decay of the spacetime itself.  The absence of supersymmetry allows one to tune the two effects completely independently.  However, in the supersymmetric context the two are tied together (in the absence of open string tachyons), and this significantly restricts the outcome.  This ability to tune arbitrarily should not be confused with bosonic branes being `more stable'.  Furthermore, as we will see in the next section, these bulk operators (marginal or relevant) can force one to include relevant boundary operators even if you set them to zero by hand before perturbing the theory.

\subsection{T-dual description}

Since we are using T-duality, let us describe the T-dual branes explicitly.  The T-dual of type II on a Scherk-Schwarz circle is a type 0 theory on a circle with a $(-1)^{F_R}$ orbifold, where $F_R$ is the right moving worldsheet fermion number \cite{Bergman:1999km, Imamura:1999um}.  The type 0 GSO projection is of the form $P^{NSNS}_{GSO}= \frac{1}{2}(1+(-1)^{F_L+F_R})$ and $P^{RR}_{GSO}= \frac{1}{2}(1\pm (-1)^{F_L+F_R})$.  In flat space, this projection results in two types of branes for a given dimensionality, one for each sign of $\eta$.  With the orbifold, there is no longer a global definition of each type of brane.  Under a rotation about the circle, a magnetic brane will turn into an electric brane.  These correspond to a change in sign of the Wilson lines on the type II wrapped branes.  Thus the superposition of a magnetic and electric brane in type 0 gives the right boundary state for type II and is invariant under the type 0 orbifold action.

Under tachyon condensation, one of the type 0 branes becomes massive and the other decays.  In the type II description this is simply a choice of the sign of the Wilson line (relative to $\eta$), which altered the form of the semi-classical boundary state.  Now, this difference can be attributed a slightly more physical description.  The tachyon makes one brane massive and causes the other to decay all while destroying the orbifold.  In this T-dual picture, the unaltered branes sit at a special point on the circle.

\section{Worldsheet RG and Boundary Perturbations}
\subsection{Bulk induced Boundary Operators}

In a more general CFT context, we are interested in the modification of the boundary theory by the presence of a relevant operator in the bulk.  There has been surprisingly little work in this area, with only recent results like those in \cite{Fredenhagen:2006dn} and references therein.  While the previous work stressed the addition of a particular boundary operator enforced by supersymmetry, we might inquire about the RG flow more generally.  At the one loop level, all the necessary information is in \cite{Fredenhagen:2006dn}.  We will adopt their language in what follows.

Let us start with a CFT and deform it by both bulk and boundary operators of the form,
\beq
S = S_{CFT} + \sum_{i} {\lambda_i l^{2-h_{\phi}} \int d^2z} \phi_{i}(z) + \sum_{j} {\mu_j l^{1-h_{\psi}}\oint dx \psi_{j}(x)}.
\eeq
Where $h_{O}$ is the dimension of the operator and $\lambda$ and $\mu$ are dimensionless couplings.  Given $\lambda_i \neq 0$, the D-branes are deformed if quantum corrections force $\mu_j \neq 0$ for some $j$.  In terms of the beta functions, this true when $\lambda_i \neq 0$, $\mu_j=0$ is not a fixed point.  The one loop corrections can be calculated easily, and was shown in \cite{Fredenhagen:2006dn} that the one loop beta functions take the form
\beq
\label{beta}
\dot{\lambda_k} = (2-h_{\phi_k})\lambda + \pi C_{ijk} \lambda_{i} \lambda_{j} + \ldots \nonumber \\
\dot{\mu_{k}} = (1-h_{\psi})\mu_k + {1 \over 2} B_{ik} \lambda_{k} + D_{ijk} \mu_{i} \mu_{j} + \ldots.
\eeq
Here $B_{ik}$, $C_{ijk}$ and $D_{ijk}$ are the coefficients for the bulk-boundary, bulk and boundary OPEs respectively.  From here, it is clear that if $B_{ik} \neq 0$ for some relevant boundary operator (given $\lambda \neq 0$), then there will be boundary RG flows induced by the bulk theory.  This is precisely what was found in \cite{Fredenhagen:2006dn} in the case of a free boson theory at the self-dual radius.

Let us now us these results for the case of a closed string tachyon (i.e. a relevant bulk operator).  This case was studied independently by \cite{Lawrence}, but we will repeat the discussion here in order to make contact the above results.  We will focus on the case of momentum or winding modes tachyons in a free theory with Dirichlet or Neumann boundary conditions.  These theories are easily described using the method of images.  In particular, if we perturb the theory by the operator
\beq
\phi \propto (e^{ik X} - e^{-ik X}),
\eeq
and impose $X=0$ on the boundary, then the bulk-boundary OPE is zero.  In other words, the boundary theory is not deformed if we impose that we sit at the zero of this operator.  However, if we take Neumann conditions on $X$, or Dirichlet conditions at a different value of $X$ then the bulk-boundary OPE will be non-zero and the boundary theory will flow away from the original fixed point.  One should note that this isn't quite what happened in the SUSY picture.  There, the bulk operator that we turned on had no zeros (as a function of $\tilde{\Omega}$).  However, SUSY imposes extra structure that allows essentially the same trick without requiring the tachyon vev to vanish.

Finally, let us consider the more general perturbed CFT.  The above statement about the survival of branes was achieved by finding conformally invariant boundary conditions that were not modified by the bulk flow.  So, let us consider an analogous situation for an arbitrary CFT.  We deform the theory by an operator $\phi$ which is relevant or marginal.  If there exist conformally invariant boundary conditions such that $\langle \phi \rangle =0$ on the disk, then two conclusions follow.  First, $B_{\phi 1} = 0$.  This is particularly useful if the bulk operator is of the form $\phi(z) = \prod{\phi_{i}}(z)$ with $\phi_{i}(z) \phi_{j}(y) \propto \delta_{ij}$.  In such a case, there are no induced boundary terms for the form $\psi_i 1_j$ if $\langle \phi_j \rangle =0$.  The second implication follows from the statement that
\beq
\langle \phi \rangle \propto \sum{B_{\phi j}}\langle \psi_j \rangle =0.
\eeq
We conclude immediately that either $B_{\phi j}=0$ or $\langle \psi_j \rangle =0$.  This does not eliminate the possibility of induced boundary flows, but does limit the types of deformations that can occur.  In the above cases, this is sufficient to eliminate the one-loop boundary flow.

\subsection{Conformal Invariance}

At the classical level, the statement that the boundary conditions are conformal is the requirement that $T=\bar{T}$ on the boundary.  This is simply the requirement that the off diagonal components of the energy-momentum tensor vanish on the real axis (if we work in the upper half plane).  This is the statement that there is no energy flow through the boundary.  Throughout this discussion, we have only insisted on classical conformal invariance of both the boundary and bulk operators.  One might ask whether it is possible to make the chosen boundary conditions conformal beyond the classical level.  

At the quantum mechanical level, conformal invariance requires the vanishing of all beta functions, including boundary operators.  The statement that our boundary conditions are exactly conformal is equivalent to saying that the beta functions associated with all boundary operators vanish.  In the absence of a tachyon, this requirement imposes the usual equations of motion, which for the boundary operators can be derived from the DBI action \cite{Leigh:1989jq}.  The addition of the tachyon should simply change the equations of motion, causing the brane to move (see \cite{Graham:2006gc} for a nice discussion of similar effects in Liouville theory).

Let us sketch the calculation in the toy CFT described in section 5.1.  The calculations are simpler and contain the relevant physics, but we will briefly comment on the difference in the SUSY case below.  We have already seen the leading term in the beta function that arises from the bulk operator.  Let us now calculate the higher order terms using the background field method, following \cite{Leigh:1989jq}.  In particular, we will choose boundary conditions such that $X_{\mu} = f_{\mu}(\zeta)$ where for simplicity we will only consider $\mu = 1,2$.  Here $f$ is some function that describes the subspace of the spacetime mapped out by the D-brane.  Here, the fact it is dependant on a single parameter $\zeta$ means that this is a D0 brane.  For example, the case of a motionless D0 at the point $X_1=0$ on the circle is given by $f_0 = \zeta$ and $f_1 = 0$.  In what follows, by definition $f^(0)_1 \equiv \Omega$, on the boundary of the worldsheet.

Additional divergences are calculated by expanding $X_{\mu} \rightarrow X_{\mu}+\xi$, and then integrating out $\xi$.  On the boundary we can then expand $\xi$ in $\zeta$ and derivatives of $f(\zeta)$.  The additional terms in the beta functions for the operator $\exp(-\kappa X_0)$ will then be proportional to the derivatives of $f$.  If the beta function cannot be cancelled by setting $f_1$ to a constant, then this indicates the force on the brane.

The boundary Langragian is given in powers $\zeta$ as $\mathcal{L}_{bdy}= \sum \mathcal{L}^{(n)}_{bdy}(\zeta^{n})$.  To second order in $\zeta$ we get
\beq
\mathcal{L}^{(1)}_{bdy} =\zeta \{f'_{\mu} \partial_{n}X^{\mu}- \eta \mu e^{-\kappa X_{0}}(\kappa f'_{0} \cos(w \Omega)+ w f'_{1} \sin(w \Omega))\}
\eeq
\beq
\mathcal{L}^{(2)}_{bdy} ={\zeta^2\over{2}} \{ f''_{\mu}\partial_{n}X^{\mu} + \eta \mu e^{-\kappa X_{0}}(((\kappa f'_{0})^2-\kappa f''_{0})\cos(w \Omega) \nonumber \\-(w f'_{1})^2 \cos(w \Omega)-w f''_{1} \sin(w \Omega))\}
\eeq
where the $f_{\mu}^{(n)}$ denotes derivates with respect to $\zeta$ and $\partial_{n}$ is the derivative normal to the boundary.  The linear piece is set to zero by imposing the boundary conditions.  These will relate the operators $\partial_{n}X$ with the boundary potential (thus they are not independent boundary operators).  One loop diagrams in $\zeta$ (when we include all the allowed boundary operators) lead to divergences in front of the operators in the quadratic Langragian.  Thus, the beta functions for $e^{-\kappa X_{0}}$ (or more accurately $\mu$) will contain contributions proportional to derivatives in $f$.

It should be clear that Dirichlet boundary conditions with $\cos(w \Omega)=0$ eliminate the divergences renormalizing $\mu$.  Furthermore, with these boundary conditions, the bulk induced piece vanishes as well.  Thus, taking modified Neumann conditions on $X_0$ and setting $f^{(n)}_{1}=0$ for all $n>0$ is sufficient to set the beta functions to zero.  In other words, this choice of boundary condition is conformal.

For other choices of the Dirichlet boundary condition, there are additional divergences which will require solving more complicated equations of motion.  However, there is no consistent approximation that lets one terminate the equations of motion at quadratic order.  In particular, loop corrections are not suppressed since the size the circle is string scale (in other words, $w$ is order 1).  Thus control is maintained only while the tachyon vev and the higher derivatives of $f$ are small.  Therefore, the approximation will break down before we reach the fixed point (in many cases).

Nevertheless, conformal invariance at higher orders in $\alpha'$ does precisely what one might expect from the interpretation given in terms of masses.  The branes start to roll to the state of lowest mass, which is a conformally invariant solution.  Further analysis is needed to ascertain whether or not the branes all roll to this point in the large tachyon background.

In the supersymmetric case, the situation is slightly different.  The boundary operator of concern is not induced by a divergent bulk-boundary correlator, but instead by worldsheet SUSY.  In particular, we can repeat background field calculations in the section the bosonic fields with chiral superfields.  The boundary term linear in the perturbation is
\beq
\mathcal{L}^{(1)}_{bdy} =\int d\tilde{\theta} \bar{\zeta}(D\bar{X}-\eta \bar{D}\bar{X})(\theta=\eta \bar{\theta}=\tilde{\theta}, z=\bar{z}) 
\eeq
where $D,\bar{D}$ are super-derivatives.  The bosonic piece after integrating over $\tilde{\theta}$ is precisely the boundary term given in (\ref{bc}).  In the case of the CFT, the bulk induced term was what required moving the D-brane.  In the absence of such a term, the first divergence arises at second order in derivatives of the tachyon vev.  Therefore, if we set $f^{(n)}_{1}=0$, then conformal invariance is broken at order $\kappa^2$.  Thus, the boundary conditions used are conformal to the level of approximation being used.

\section{Discussion}

D-branes have been studied under closed string tachyon condensation previously in the context of gauged linear sigma models \cite{Minwalla:2003hj,Martinec:2002wg, Moore:2004yt, Moore:2005wp}.  Linear sigma models have been useful in the study of closed string tachyons largely because it provides a framework where both endpoints can be described from a single set of operators.  Studying branes in these models can be done by adding a boundary to the GLSM and following the boundary conditions/operators under the RG flow associated with the tachyon condensation.  While this is has been a successful way to understand the evolution of branes into other branes, it will not provide any information about branes in the condensate itself.  This is of particular importance when the condensation is localized in time but not space.

In \cite{Adams:2005rb}, in the sigma model description, extra vacua appeared in the final state.  One might ask if the unmodified branes are actually ending in the extra vacua.  In \cite{McGreevy:2005ci}, it was speculated that these extra vacua were related to the sign of the scalar potential in the Lorentzian description (the F term is strictly positive in the Euclidean description of both the Heterotic and type II).  Both the Wilson line and the sign of the Lorentzian potential is a function of the T-dual coordinate.  If the branes of interest are those that lie at $\cos(w \tilde{\Omega})=0$, then these are explicitly in at the positive values of the potential.  Nevertheless, it is not impossible that there is a connection between the two.

The branes of interest in the nothing state are equivalent to the D2 branes in the GLSM.  The Wilson lines on these branes are generated from boundary fields coupling to bulk fields.  Meanwhile, the dust vacua arise from the $\sigma$ vacua of GLSM.  This model has $\sum q_i < 0$, which drives the FI term $\rho>0$ in the IR.  We will interpret the change in topology as the appearance of the `nothing' state.  The appropriate boundary conditions on the D2 branes is $\Sigma=\bar{\Sigma}$.  The $\sigma$ vacua may not necessarily take real values, but there do exist solutions with this property.  The role of these additional vacua in GLSMs has been studied in \cite{Martinec:2002wg, Moore:2004yt, Moore:2005wp, Melnikov:2005hq} including information about D-brane charges.  It remains unclear what relationship the branes in these vacua have to the branes discussed above from a worldsheet analysis.  Understanding relationship between the branes might also clear up the nature of the relationship between the sign of the (Lorentzian) tachyon potential and the $\sigma$ vacua in the GLSM.

There remain many additional open questions.  The most obvious being that of the open strings.  In principal, there are worldsheet techniques that might make it possible to extract interesting information, but the work is ongoing.  In general, the results presented here and elsewhere from worldsheet and GLSM approaches provide some interesting clues.  However, it is clear that these techniques are limited in the types of information that can be retrieved.  AdS/CFT is the only non-perturbative definition of the theory known that might improve this understanding, but it remains to be seen if we can find an appropriate context there to study this type of situation.

\acknowledgments
I would like to thank David Starr for early collaboration and many helpful discussions.  I would also like to thank Albion Lawrence, Xiao Liu, John McGreevy and Eva Silverstein for many helpful discussions.  I would also like to thank the KITP for their hospitality during a portion of this work.  This project was supported in part by a NSERC Postgraduate fellowship, by the DOE under contract DE-AC03-76SF00515 and by the NSF under contract 9870115 and under Grant No. PHY99-07949.

\end{document}